\documentclass{edp-conf}
\usepackage{graphicx}
\begin{document}

\TitreGlobal{SF2A 2005}

%%-----------------------------
%%      the top matter
%%-----------------------------
\title{Features of radio-detected Extensive Air Shower with CODALEMA}
\author{D. Ardouin$^1$}
\author{{\underline {A. Bell\'etoile$^1$}}}
\author{D. Charrier$^1$}
\author{R. Dallier}\address{SUBATECH, In2p3-CNRS/Universit\'e de Nantes/Ecole des Mines de Nantes}
\author{L. Denis}\address{Station de Radioastronomie de Nan\c{c}ay}
\author{P. Eschstruth}\address{LAL, In2p3-CNRS/Universit\'e de Paris Sud, Orsay}
\author{T. Gousset$^1$}
\author{F. Haddad$^1$}
\author{P. Lautridou$^1$}
\author{A. Lecacheux}\address{LESIA, Observatoire de Paris-Meudon}
\author{D. Monnier-Ragaigne$^3$}
\author{O. Ravel$^1$}
\runningtitle{Radio-detection of Extensive Air Showers}
\setcounter{page}{1}
% Keep this line, even pl_SF2A2005if the page will be settled afterwards..

\index{Ardouin, D.}
\index{Bell\'etoile, A.}
\index{Charrier, D.}
\index{Dallier, R.}
\index{Denis, L.}
\index{Eschstruth, P.}
\index{Gousset, T.}
\index{Haddad, F.}
\index{Lautridou, P.}
\index{Lecacheux, A.}
\index{Monnier-Ragaigne, D.}
\index{Ravel, O.}
% Repeat the authors here, this will help to make the final index

\maketitle
\begin{abstract} 
Some performances of the present CODALEMA experiment, set up to analyse radio-detected Extensive Air Shower (EAS) events, are presented. Characteristics of the EAS electric field distribution sampled on a 600~m long axis are discussed.
\end{abstract}
%
%%-----------------------------
%%      your text
%%-----------------------------
Based on an idea of detection put forward in the 60's (Askar'yan 1962, Weekes 2001, Allan 1971), the CODALEMA experiment aims to study cosmic rays via transient signals in the radio frequency domain 1-100~MHz. This device has already shown its ability to sign radio emission counterparts of air showers with an energy threshold of $5.10^{16}$~eV (Ardouin et al. 2005~a-b-c).
\section{ The CODALEMA experiment}

\begin{figure}
\begin{center}
\includegraphics[width=10cm, height=4cm]{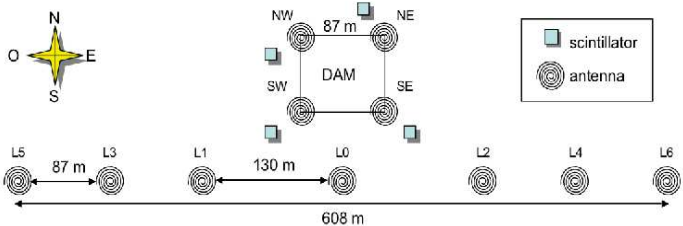}
\end{center}
\caption{Current CODALEMA setup. The particle detectors (scintillators) act as a trigger with a fourfold coincidence requirement.}
\label{fig:setup2}
\end{figure}

The set-up (Fig.~\ref{fig:setup2}) is made up of 11 log periodic antennas~(http://www.obs-nancay.fr) of the decametric array of Nan\c cay and 4 particle detectors originally designed as prototypes for the Pierre Auger Observatory~(Boratav et al. 1995).

The experiment is triggered out when a four particle detector coincidence occurs within 600 ns. Considering an acceptance of $16.10^{3}$~m$^{2}\times$sr, the trigger counting rate of 0.7 event per minute leads to an energy threshold of $1.10^{15}$~eV. For each trigger, antenna signals are recorded after RF amplification (1-200~MHz, gain 35~dB) and analog filtering (24-82~MHz) as a time function by 8 bits ADCs (500~MHz sampling, 10~$\mu$s recording time).

Effective EAS radio events, that represent only a fraction of the total amount of recorded data, are discriminated by an off-line analysis: radio signals are first numerically filtered (37-70~MHz) to detect radio transients. If the amplitude threshold condition, based on a noise level estimation, is fulfilled, the arrival time of the electric wave is set at the point of maximum squared voltage of the transient. When at least 3 antennas are fired, the arrival direction of the radio wave is calculated using a plane front fit. Through the arrival time distribution between radio wave and particle fronts, fortuitous events due to anthropic emitters and atmospheric phenomenon with a flat distribution and the EAS candidates which have a sharp peak distribution of a few tens of nanoseconds are identified. Within this peak, the true radio-particle coincidences are finally selected using a 15 degrees cut in the distribution of the angular differences between the arrival directions reconstructed from antenna signals and from scintillators (Ardouin et al. 2005~b).

At the end of this procedure, the resulting counting rate of EAS events with a radio contribution is 1 per day. Assuming in a first approximation that acceptances of both the antenna and the particle detector arrays are the same, an energy threshold of about $5.10^{16}$~eV is deduced for the radio detection~(Ardouin et al. 2005~a).

\section{Characteristics of the radio-EAS events}

Each CODALEMA antenna allows to associate a measured electric field value to its location. This amounts to a mapping of the electric field. In case of a radio EAS event, the Electric Field Profile (EFP) can be obtained by fitting a decreasing exponential, given Eq.~\ref{eq:fit}~(Allan 1971).

\begin{equation}
E(d) = E_0.\exp[\frac{-d}{d_0}].\
\label{eq:fit}
\end{equation}

\begin{figure}
\begin{center}
\includegraphics[width=5cm, height=3.5cm]{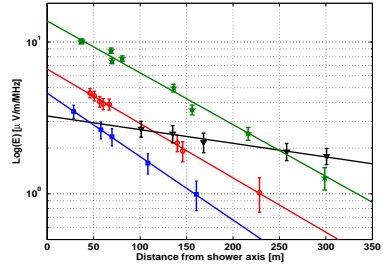}
\end{center}
\caption{Electric Field Profile (EFP) of a set of radio EAS events recorded on CODALEMA. The measured amplitude in $\mu$V/m/MHz is plotted versus $d$, the distance from the antenna to the shower axis in meters with an error deduced from the noise estimated on each event. Parameters are the shower core position, $E_0$ and $d_0$ from Eq.\ref{eq:fit}.}
\label{fig:EFP}
\end{figure}

First, the core position of the shower is roughly estimated by a barycenter calculation~(Ardouin et al. 2005 c) of the field amplitude on both North-South and East-West axis of the array. Then, using this core location estimation and the reconstructed arrival direction of the shower, the measured voltage on each tagged antenna is projected in a shower based coordinate system. Finally, the EFP function is fitted with the shower core position and both $E_0$ and $d_0$ as free parameters. A sub-set of 4 events is shown Fig.~\ref{fig:EFP} as an illustration. It has been then computed on a set of 60 events with a sufficient antenna multiplicity and a good minimisation was obtained.

\begin{figure}[h]
\begin{center}
\includegraphics[width=5cm]{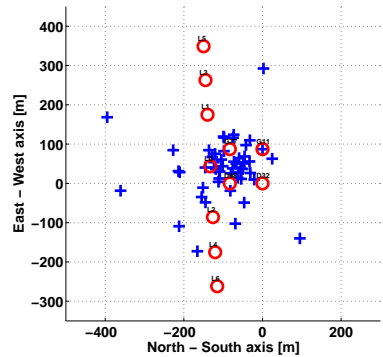}
\end{center}
\caption{ Fitted core locations of 60 EAS events (crosses) plotted with respect to the CODALEMA setup (circles are antennas) on both North-South and West-East axis.}
\label{fig:PARAM}
\end{figure}

Due to the nature of the trigger (particle detectors), our system mainly detects showers falling in the vicinity of the array~(Ardouin et al. 2005~a). The fitted shower core positions are plotted Fig.~\ref{fig:PARAM}, in the CODALEMA setup coordinate system.

Considering the obvious lack of statistics, preliminarly analysis shows no simple relation to be clearly identified yet. Nevertheless, fitted core positions appear to be variable on a reasonnable scale just like $E_0$, which is spread from a few $\mu$V/m/MHZ to some 25~$\mu$V/m/MHz and $d_0$ that goes from 100~m which is approximatively the pitch of the array to more than 300~m. A dependence of the electric field amplitude parameter $E_0$ on the energy of the primary particle~(Huege and Falcke 2005) has been predicted but a calibration of the instrument is first needed. This operation is already beeing conducted by adding more particle detectors to the CODALEMA array. In the same way, we expect the slope of the exponential fit to be related to the zenithal angle of the shower and then to the primary particle nature and energy. Again, the calibration of the system and a larger amount of data will offer the possiblity to identify each physical contribution.

\section{Conclusion}

Electric field transients generated by extensive air showers have been measured with CODALEMA. The current effective counting rate of 1 event a day leads to a statistical energy threshold around $5.10^{16}$~eV. Shower core location of an EAS can be determined on an event by event basis using the Electric Field Profile function. The EFP function could also constitue a purely ``Radio'' discrimination criterion, which would be one further step towards a stand-alone system. Investigations are also currently lead on the feasibility of adding radio detection techniques to an existing surface detector such as the Pierre Auger Observatory~(Auger Collaboration 1999) in order to reach a higher energy range. In the future, we expect that the radio signals could provide complementary information about the longitudinal development of the shower, as well as the ability to lower the energy threshold.
%%-----------------------------
%%      your bibliography
%%-----------------------------
%In preparing the reference list please adhere to the following format.
% Attention should be paid to the order of the items in each reference
% and to the punctuation used. Please see Sect. 4 in the User's Guide
% that comes with the new macro package.

\end{document}